\documentclass[12pt]{article}
\usepackage{latexsym}
\usepackage{amsmath}
\renewcommand{\epsilon}{\varepsilon}
\renewcommand{\arraystretch}{1.15}

\usepackage{amsmath,bm}
\usepackage{amssymb}
\usepackage{xcolor}
\usepackage{amsthm}
\usepackage[ansinew]{inputenc}
\usepackage{graphicx}
\usepackage{epsfig}
\usepackage{dsfont}
\usepackage{bbm}
\usepackage{url}
\usepackage{placeins}
\usepackage{subfigure} 
\usepackage{multirow}

\usepackage[authoryear]{natbib}

\newtheorem{satz}{Theorem}[section]

\newtheorem{algorithm}[satz]{Algorithm}

\def\3{\ss}

\newcommand{\bea}{\begin{eqnarray*}}
	\newcommand{\eea}{\end{eqnarray*}}
\newcommand{\be}{\begin{eqnarray}}
\newcommand{\ee}{\end{eqnarray}}

\newcommand{\ba}{\begin{array}}
	\newcommand{\ea}{\end{array}}

\def\3{\ss}

\usepackage[authoryear]{natbib}
\usepackage{subfigure} 

\begin{document}
	
	\title{{\bf Equivalence tests for binary efficacy-toxicity responses}}

	\author{
	Kathrin M\"ollenhoff$^{1}$,
	Holger Dette$^{1}$  ,
	Frank Bretz$^{2}$ 	\bigskip\\
	\small $^{1}$	Department of Mathematics, Ruhr-Universit\"at Bochum, Germany,\\
	\small $^{2}$Novartis Pharma AG, CH-4002 Basel, Switzerland,\\
	}

	\pdfminorversion=4
	\maketitle
	
	\begin{abstract}
	
Clinical trials often aim to compare a new drug with a reference treatment in terms of efficacy and/or toxicity depending on covariates such as, for example, the dose level of the drug. Equivalence of these treatments can be claimed if the difference in average outcome is below a certain threshold over the covariate range. In this paper we assume that the efficacy and toxicity of the treatments are measured as binary outcome variables and we address two problems. First, we develop a new test procedure for the assessment of equivalence of two treatments over the entire covariate range for a single binary endpoint. Our approach is based on a parametric bootstrap, which generates data under the constraint that the distance between the  curves is equal to the pre-specified equivalence threshold. Second, we address equivalence for bivariate binary (correlated) outcomes by extending the previous approach for a univariate response. For this purpose we use  a $2$-dimensional Gumbel model for binary efficacy-toxicity responses.  We investigate the operating characteristics of the proposed approaches by means of a simulation study and present a case study as an illustration.

\end{abstract}


\parindent 0cm

\maketitle

\section{Introduction}
\label{sec:intro}
\def\theequation{1.\arabic{equation}}
\setcounter{equation}{0}

Equivalence tests are used in clinical drug development to assess similarity of a test treatment with a reference treatment. Considering continuous data with covariates, the effect of a drug is 
described as a function of the covariates and equivalence is claimed is these function are in some sense similar. Several authors use confidence bands   for the difference between the response functions to construct equivalence tests (see, for example \cite{liubrehaywynn2009,gsteiger2011,bretzmoelldette2016}). Alternatively, \cite{detmolvolbre2015,moellenhoff2018} proposed 
more powerful tests  by estimating  a  distance between the two functions, such as the the squared integral of the difference or the maximal deviation between the curves. They claim equivalence if the estimated distance is small. All these approaches assume continuous outcomes.
However, in some situations drug efficacy is measured using a  binary outcome (for some examples see \cite{chowliu1992,cox2018}). For example, a patient is considered to be a responder, that is the efficacy response is $1$, if the drug effect is as desired. This can be for example the shrinkage of a tumor or the curing of any disease. 
Equivalence tests have been proposed in these settings by, for example, \cite{nam1997} and \cite{chen2000}, who derive methodology for comparing the treatments in response probabilities. These authors investigate different types of test statistics and perform sample size determination in several situations  but they do not include any covariates such as, for example, the dose. 

Many clinical trials involve the measurement of a second endpoint (e.g. to assess toxicity) and hence bivariate outcomes are considered which are likely to be correlated, see for example \cite{murtaugh1990,heise1996,thall2004,dragalin2006} and \cite{gaydos2006}. This is, for instance, the case when observing efficacy and toxicity of a drug. The toxicity response is $1$ if a side-effect (e.g. fatigue or nausea) is observed.
Several methods for modelling multivariate binary outcomes have been proposed, see for example, \cite{glonek1995}. Considering efficacy-toxicity responses,  \cite{murtaugh1990} and \cite{heise1996} investigate bivariate binary responses and derive optimal designs for this situation by fitting the data to a bivariate logistic model and a Cox model (see also \cite{cox2018}). \cite{deldossi2019} propose Copula functions to model these types of outcomes. Further,  \cite{thall2004} and \cite{dragalin2006} develop adaptive designs for identifying the optimal safe dose.
Finally several authors investigate the modeling and design of phase I/II dose-finding trials incorporating bivariate outcomes using Bayesian methods, see for instance \cite{nebiyou2005,zhang2006,yin2006}.

Different to the literature reviewed above, we investigate equivalence tests with The purpose to assess similarity of a  reference and a test treatment for efficacy and toxicity. Equivalence can only be claimed if the differences of \textit{both} outcomes are below prespecified thresholds over the complete range of covariates. 
Accordingly, we first develop a new test for assessing equivalence in case of a single binary endpoint over the range of covariates. Second, we address equivalence for bivariate binary (correlated) outcomes and develop  an equivalence  test for comparing simultaneously efficacy and toxicity of a reference to a test treatment. For this purpose we use  a $2$-dimensional Gumbel model (see \cite{gumbel1961}) for bivariate logistic regression  to model correlated bivariate binary endpoints.   Our approach is based on a non-standard  parametric bootstrap, which generates data under the constraint that the distances between the  curves are precisely equal to the thresholds.
We investigate finite sample properties and illustrate the procedures with a clinical trial example.

\section{Comparing curves for binary outcomes}
\label{methods}
\def\theequation{2.\arabic{equation}}
\setcounter{equation}{0}

In this section we introduce a model-based approach for the investigation of equivalence between the efficacy of two treatments assuming binary endpoints. 
We consider models with covariates and assume for simplicity a one-dimensional covariate, although the proposed methodology applies more broadly.  For both treatment groups we choose the covariate space as a (log-transformed) dose range $\cal{D}$ and assume that treatments are conducted at  $k_\ell$ dose levels $d_{\ell,1},\ldots,d_{\ell,k_\ell}$, $\ell=1,2$, where  the index $\ell=1$ corresponds to the reference and $\ell=2$ to the test treatment. At dose level $d_{\ell,i}$ we observe $n_{\ell,i}$ patients, $i=1,\ldots,k_\ell$. Let $Y_{\ell,i,j}$ denote the binary outcome for the $j$th patient allocated to the $i$th dose level receiving treatment $\ell$.  If a patient responds to the drug, we have $Y_{\ell,i,j}=1$, otherwise $Y_{\ell,i,j}=0$. More precisely,  $Y_{\ell,i,j}$ follows a Bernoulli distribution with parameter $p_\ell(d_{\ell,i})$ modelling the probability of success under treatment $\ell$ with dose level $d_{\ell,i}$, $i=1,\ldots k_\ell$, $\ell=1,2$.
We use regression techniques to model the dose-response relationship. More precisely,  the probability of the $j$th patient allocated to dose level $d_{\ell,i}$ responding to treatment $\ell$ is given by 
\be\label{log_reg}
p_\ell(d_{\ell,i})=
\mathbb{P}(Y_{\ell,i,j}=1 \mid d_{\ell,i}  )=  
\eta^E_{\ell}(d_{\ell,i},\beta_\ell,\gamma_\ell) ,\ \ell=1,2,
\ee
where $\eta^E_{\ell}$ is a known distribution function determined by
parameters $ \beta_\ell,\gamma_\ell$. Hence the curve $\eta^E_{\ell}(d,\beta_\ell,\gamma_\ell)$ models the probability of efficacy over the entire dose range.  

Common examples of \eqref{log_reg} include the logistic regression model $\mathbb{P}(Y_{\ell,i,j}=1  \mid d_{\ell,i})= \frac{1}{1+e^{-\beta_\ell-\gamma_\ell \cdot  d_{\ell,i}}}$ and the probit regression model $\mathbb{P}(Y_{\ell,i,j}=1  \mid d_{\ell,i})= \Phi(\beta_\ell+\gamma_\ell \cdot d_{\ell,i} ),\ \ell=1,2$, where $\Phi$ is the distribution function of the standard normal distribution (see for example \cite{long2006}).
Assuming independent observations, the likelihood of the observed data in treatment group $\ell=1,2$ is

\begin{align*}
\mathcal{L}_\ell (\beta_\ell,\gamma_\ell|y_{\ell,1,1},\ldots,y_{\ell,k_\ell,1},\ldots,y_{\ell,k_\ell,n_{\ell,k_\ell}})  &=\prod_{i=1}^{k_\ell}\prod_{j=1}^{n_{\ell,i}} p_\ell(d_{\ell,i})^{y_{\ell,i,j}} (1-p_\ell(d_{\ell,i}))^{(1-y_{\ell,i,j})}  \nonumber\\
&=\prod_{i=1}^{k_\ell }p_\ell(d_{\ell,i})^{\zeta_{\ell,i}}(1-p_\ell(d_{\ell,i}))^{n_{\ell,i}-\zeta_{\ell,i}},
\end{align*}
where  $\zeta_{\ell,i}:=\sum_{j=1}^{n_{\ell,i}}y_{\ell,i,j}, \ i=1,\ldots,k_\ell,\ \ell=1,2$. Taking the logarithm yields
\begin{align}\label{mle-univ}
l_\ell(\beta_\ell,\gamma_\ell)&:=\log{\mathcal{L}(\beta_\ell,\gamma_\ell|y_{1,1},\ldots,y_{k,1},\ldots,y_{k,n_k})}\nonumber\\
&=\sum_{i=1}^{k_\ell} \zeta_{\ell,i}\log{p_\ell(d_{\ell,i})}+(n_{\ell,i}-\zeta_{\ell,i})\log{(1-p_\ell(d_{\ell,i}))}
\end{align}
and corresponding Maximum-Likelihood-estimates (MLE) are obtained by maximizing the function \eqref{mle-univ}.

In order to investigate the difference in efficacy between the reference and the test treatment we consider the maximal deviation between the two curves in \eqref{log_reg} and define the equivalence hypotheses by
 \begin{equation} \label{hypotheses-univ} 
	H_0^E:  \max_{d\in \cal D}\left| \eta^E_1(d,\beta_{1},\gamma_{1})-\eta^E_2(d,\beta_{2},\gamma_{2})\right| \geq \epsilon^E \text{ vs. } H_1^E:\max_{d\in \cal D}\left| \eta^E_1(d,\beta_{1},\gamma_{1})-\eta^E_2(d,\beta_{2},\gamma_{2})\right| < \epsilon^E, \end{equation}
where $\epsilon^E$ denotes a prespecified margin of equivalence in efficacy between the two curves, which has to be carefully chosen by clinicians in advance. 
The choice of these thresholds is a common issue and there are no general recommendations available. However, according to guidelines (\cite{food2003guidance}) equivalence margins between $0.1$ and $0.2$, that is a deviation of the two products between $10\%$ to $20\%$, seem to be reasonable.

The following algorithm provides a bootstrap test for the hypotheses \eqref{hypotheses-univ}, which keeps its nominal level, say $\alpha$, and is consistent. It is derived by adapting the methodology developed in  \cite{detmolvolbre2015} to binary data.

\begin{algorithm}
	\label{alg1} {\rm (parametric bootstrap for testing equivalence of binary outcomes)
		\begin{itemize}
			\item[(1)]
			Calculate the MLE $(\hat\beta_{\ell},\hat\gamma_{\ell})$, $\ell=1,2$, by maximizing for each group the log-likelihood given in \eqref{mle-univ}.
			The test statistic is obtained by
			$$\hat d^E:=\max_{d\in \cal D}\left|  \eta^E_1(d,\hat\beta_{1},\hat\gamma_{1})-\eta^E_2(d,\hat\beta_{2},\hat\gamma_{2})\right|.$$
			\item[(2)]  
			Define estimators of the parameters $\beta_\ell, \gamma_\ell,\ \ell=1,2$, so that the corresponding curves fulfill the null hypothesis \eqref{hypotheses-univ}, that is
			\begin{equation} \label{MLcons}
			\big(\hat{\hat{\beta}}_{\ell},\hat{\hat{\gamma}}_{\ell}\big)= \left\{
			\begin{array} {ccc}
			(\hat\beta_{\ell},\hat\gamma_{\ell}) & \mbox{if} & \hat d^E \geq \varepsilon \nonumber \\
		   (\bar\beta_{\ell},\bar\gamma_{\ell})& \mbox{if} & \hat d^E < \varepsilon
			\end{array}  \right. \quad \ell=1,2,
			\end{equation}
			where $(\bar\beta_1,\bar\gamma_{1})$ and $(\bar\beta_{2},\bar\gamma_{2})$ minimize the same objective function as defined in \eqref{mle-univ}, but 
			under the constraint
			\begin{equation}\label{constr}
			d^E=\max_{d\in \mathcal{D}} \left| \eta^E_1(d,\beta_{1},\gamma_{1})-\eta^E_2(d,\beta_{2},\gamma_{2})\right|= \epsilon^E.
			\end{equation}
 We discretize the dose range ${\cal D}$ to get a feasible optimization problem by fixing $r$ nodes $d_1,\dots,d_r$ and use the smooth approximation (as $ \lambda \to 0$)
  $$
\max_{i=1}^r d_i \approx 
\lambda\log{\sum_{i=1}^r\exp{\tfrac{d_i}{\lambda}}}
$$ 
for the calculation of the maximum in \eqref{constr}.
Finally the optimization procedure is performed by running the auglag() function implemented in the R package alabama by \cite{alabama}. The algorithm implemented in this function is based on the augmented Lagrangian minimization algorithm, which is typically used for solving constrained optimization problems.
				\item[(3)]
				Proceed as follows:
			\begin{itemize}
				\item[(i)] 	 Generate bootstrap data under the null hypothesis \eqref{hypotheses-univ}, that is, create binary data specified by the parameters $\big(\hat{\hat{\beta}}_{\ell},\hat{\hat{\gamma}}_{\ell}\big)$, $\ell=1,2$. 
				More precisely, calculate $\eta^E_\ell(d_{\ell,i},\hat{\hat{\beta}}_{\ell},\hat{\hat{\gamma}}_{\ell})$, $i=1,\ldots,k_\ell,\ \ell=1,2$ yielding the probabilities of success $p(d_{\ell,i})$ at each dose level $d_{\ell,i} $. 
				\item[(ii)] From the bootstrap data calculate  the MLE  $(\hat\beta_{\ell}^*,\hat\gamma_{\ell}^*)$ as in step $(1)$ and the test statistic
				\be\label{boot}\hat d^{E*}=\max_{d\in \cal D}\left| \eta^E_1(d,\hat\beta^*_{1},\hat\gamma^*_{1})-\eta^E_2(d,\hat\beta^*_{2},\hat\gamma^*_{2}).\right|\ee
		\item[(iii)]	Repeat the steps (i) and (ii) $n_{boot}$ times to generate  replicates $\hat d^{E*}_{\infty,1}, \dots, \hat d^{E*}_{\infty,n_{boot}}$ of $\hat d^{E*}$.    Let $\hat d^{E*(1)} \le \ldots  \le \hat d^{E*(n_{boot})}$
			denote the corresponding order statistic. The estimator of the $\alpha$-quantile of the distribution of $\hat d^{*}$
			is defined by  $\hat d^{E*(\lfloor n_{boot} \alpha \rfloor )}$. Reject the null hypothesis \eqref{hypotheses-univ}, if
			\begin{equation} \label{testInf}
			\hat d^E < \hat d^{E*(\lfloor n_{boot} \alpha \rfloor )}.
			\end{equation}
			Further we obtain the $p$-value by $\hat F_{n_{boot}}(\hat d^E)$, where $\hat F_{n_{boot}}(x) = {\frac{1}{n_{boot}}} \sum_{i=1}^{n_{boot}}  I\{ \hat d^{E*}_{\infty,i} \leq x\} $ 
 denotes the empirical distribution function of the bootstrap sample. 
	\end{itemize}			\end{itemize}
}
\end{algorithm}
Note that the bootstrap quantile $\hat d^{E*(\lfloor n_{boot} \alpha \rfloor )}$ depends on the number of bootstrap replicates $n_{boot}$ and the threshold $\varepsilon^E$  given in
the hypotheses \eqref{hypotheses-univ}, but we do not reflect this dependence in our notation.
The test proposed in Algorithm \ref{alg1} has asymptotic level $\alpha$ and is consistent. More precisely,
note that $\hat d^{E*(\lfloor n_{boot} \alpha \rfloor )} \rightarrow \hat{q}_{\alpha}$ as $n_{boot}\rightarrow \infty$, where $\hat{q}_{\alpha}$ denotes the $\alpha$-quantile of the distribution of the statistic \eqref{boot}.
It  can then  be shown that under $H_0^E$
\be \label{alta}
\limsup_{n_1,n_2\rightarrow\infty} \mathbb{P}_{H_0^E}(	\hat d^E < \hat{q}_{\alpha})\leq \alpha
\ee
and that under $H_1^E$
\be \label{alt} 
\lim_{n_1,n_2\rightarrow\infty} \mathbb{P}_{H_1^E}(	\hat d^E < \hat{q}_{\alpha})=1 .
\ee
These results follow from the well-known fact that under suitable conditions of regularity the MLE converge weakly to a normal distribution (see \cite{bradley1962}), that is
\be\label{MLE_norm}
\sqrt{n_\ell}\left( (\hat\beta_{\ell},\hat\gamma_{\ell})-(\beta_\ell,\gamma_\ell)\right) \stackrel{\mathcal{D}}{\longrightarrow} \mathcal{N}(0,I^{-1}_\ell),\  \ell=1,2,
\ee
where the asymptotic variance-covariance matrix $I_\ell$ is the Fisher Information Matrix corresponding to treatment group $\ell$. The weak convergence \eqref{MLE_norm} is the essential ingredient to apply the  proof of \cite{detmolvolbre2015} to the situation considered in this paper and  \eqref{alta} and \eqref{alt} follow.

\section{Equivalence tests for efficacy-toxicity responses}\label{sec:bivariate}

\def\theequation{3.\arabic{equation}}
\setcounter{equation}{0}
\subsection{The Gumbel model for efficacy-toxicity outcomes}\label{sec:gumbel}

In this section we extend the approach of Section \ref{methods} to equivalence tests for correlated bivariate binary outcomes. 
We consider the bivariate Gumbel model (see for example \cite{murtaugh1990,heise1996}) based on the bivariate logistic function derived by \cite{gumbel1961}, which is given by
\be\label{gumbel_df} F_{U,V}(u,v)=\frac{1}{1+e^{-u}}\frac{1}{1+e^{-v}}\cdot \left(1+\frac{\nu e^{-u-v}}{(1+e^{-u})(1+e^{-v})}\right).\ee
 Note that the marginal distributions are logistic and that the parameter $\nu\in(-1,1)$ represents the dependence of $U$ and $V$. In particular the case $\nu=0$ corresponds to independent margins and in this case two separate logistic models for efficacy and toxicity can be fitted separately to the data. \\
We make the same assumptions as in the univariate case and further let $Y=(Y^E, Y^T)\in \left\{ 0,1 \right\}^2$ denote the bivariate outcome for a  patient allocated to the 
dose level $d$, where $Y^E$ denotes the efficacy and  $Y^T$ the toxicity response. We follow \cite{murtaugh1990} and formulate the model by deriving the four cell probabilities
\begin{align}\label{cellp}
p_{00}(d) :=\mathbb{P}(Y^E=0, Y^T=0|\ d)&= 1- \tfrac{1}{1+e^{-u_1(d)}}-\tfrac{1}{1+e^{-u_2(d)}}+ \tfrac{1}{1+e^{-u_1(d)}}\tfrac{1}{1+e^{-u_2(d)}}  \nonumber \\
&+ \frac{\nu e^{-u_1(d)-u_2(d)}}{(1+e^{-u_1(d)})^2(1+e^{-u_2(d)})^2}, \nonumber\\
p_{01}(d) := \mathbb{P}(Y^E=0, Y^T=1|\ d) &=  \tfrac{1}{1+e^{-u_2(d)}}- \tfrac{1}{1+e^{-u_1(d)}}\tfrac{1}{1+e^{-u_2(d)}} - \tfrac{\nu e^{-u_1(d)-u_2(d)}}{(1+e^{-u_1(d)})^2(1+e^{-u_2(d)})^2},\nonumber \\
p_{10}(d) := \mathbb{P}(Y^E=1, Y^T=0|\ d) &= \tfrac{1}{1+e^{-u_1(d)}}- \tfrac{1}{1+e^{-u_1(d)}}\tfrac{1}{1+e^{-u_2(d)}} - \tfrac{\nu e^{-u_1(d)-u_2(d)}}{(1+e^{-u_1(d)})^2(1+e^{-u_2(d)})^2},  \nonumber\\
p_{11}(d) :=\mathbb{P}(Y^E=1, Y^T=1|\ d) &=  \tfrac{1}{1+e^{-u_1(d)}}\tfrac{1}{1+e^{-u_2(d)}} + \tfrac{\nu e^{-u_1(d)-u_2(d)}}{(1+e^{-u_1(d)})^2(1+e^{-u_2(d)})^2}.
\end{align}
Here, $u_1(d)=\beta_1+\gamma_1 \cdot d$ and $u_2(d)=\beta_2+\gamma_2 \cdot d$ denote the transformed doses for efficacy and toxicity, respectively (see \cite{heise1996}). Consequently, the Gumbel model is determined by the $5$-dimensional parameter $\theta:=(\beta_1,\gamma_1,\beta_2,\gamma_2,\nu)\in \mathbb{R}^5$. The individual curves for efficacy and toxicity are obtained by the marginal probabilities
\begin{eqnarray} \label{marg}
\eta^E(d,\theta)&:=&\mathbb{P}(Y^E=1|\ d)=p_{11}(d)+p_{10}(d)=\frac{1}{1+e^{-u_1(d)}}, \nonumber\\
\eta^T(d,\theta)&:=&\mathbb{P}(Y^T=1|\ d)=p_{11}(d)+p_{01}(d)=\frac{1}{1+e^{-u_2(d)}}.
\label{ecurv}\end{eqnarray}
 Note that for simplicity we do not display the dependence on $\theta$ in the cell probability functions \eqref{cellp}. We further denote by $\eta(d,\theta):=\big(\eta^E(d,\theta),\eta^T(d,\theta)\big)$ the vector of bivariate response probabilities at dose $d$. Note that the correlation parameter $\nu$ is part of the model but not displayed explicitly. We also note that the restrictions on $\nu$ depend on the other model parameters $\beta_1,\gamma_1,\beta_2,\gamma_2$ such that all cell probabilities in \eqref{cellp} vary between $0$ and $1$ for all doses $d\in\cal{D}$. Because the correlation of $Y^E$ and $Y^T$ is given by
\be\label{corr_gumb}
corr(Y^E, Y^T|\ d)=\frac{\nu}{(e^{u_1(d) /2}+e^{-u_1(d) /2})(e^{u_2(d) /2}+e^{-u_2(d) /2})}
\ee
the upper bound of $\nu$ is at most $4$.

For the estimation of the model parameters we use again MLE. 
Therefore the likelihood for one observation $y=(y^E, y^T)\in \left\{ 0,1 \right\}^2$ modelled by the Gumbel model is given by
\begin{align}\label{mlo}
 \mathcal{L} (\theta|y)  &= p_{11}(d)^{y^E y^T} p_{01}(d)^{(1-y^E)y^T} p_{10}(d)^{y^E(1-y^T)}p_{00}(d)^{(1-y^E)(1-y^T)}.
\end{align}

\subsection{The test procedure}\label{sec:test}

Now assume that we have two groups with bivariate (efficacy/ toxicity) outcomes corresponding to the new ($\ell=1$) and reference ($\ell=2$) treatment and 
we want to compare two treatment groups with respect to their efficacy and toxicity response. 

Let $Y_{\ell,i,j}=(Y_{\ell,i,j}^E, Y_{\ell,i,j}^T)\in \left\{ 0,1 \right\}^2$ denote the bivariate outcome for the $j$th patient allocated to the $i$th dose level $d_{\ell,i}$ of treatment group $\ell$.
We observe  the data $Y_{\ell, i,j}=(Y_{\ell, i,j}^E, Y_{\ell, i,j}^T)$ and 
 denote by 
 $$\zeta^{\ell,i}_{pq}:=\sum_{j=1}^{n_{\ell,i}}I\{(y_{\ell ij}^E,y_{\ell ij}^T)=(p,q)\}$$ the number of responses with outcome $(p,q)$ at dose level $d_{\ell,i}$ in group $\ell=1,2,\ i=1,\ldots,k_\ell$. We use the Gumble model as  introduced  in Section \ref{sec:gumbel}. According to \eqref{mlo} the likelihood of the Gumbel model for group $\ell$ is given by
\begin{align*}
 &\mathcal{L}_\ell (\theta_\ell|y_{\ell,1,1},\ldots,y_{\ell,1,n_{\ell,1}},\ldots,y_{\ell,k_\ell,n_{\ell,k_\ell}})  \nonumber\\
 & = \prod_{i=1}^{k_\ell}\prod_{j=1}^{n_{\ell,i}} p_{11}(d_{\ell,i})^{y^E_{\ell ij}y^T_{\ell ij}} p_{01}(d_{\ell,i})^{(1-y^E_{\ell ij})y^T_{\ell ij}} p_{10}(d_{\ell,i})^{y^E_{\ell ij}(1-y^T_{\ell ij})}p_{00}(d_{\ell,i})^{(1-y^E_{\ell ij})(1-y^T_{\ell ij})}  \nonumber\\ 
& = \prod_{i=1}^{k_\ell} p_{11}(d_{\ell,i})^{\zeta^{\ell,i}_{11}}p_{01}(d_{\ell,i})^{\zeta^{\ell,i}_{10}}p_{10}(d_{\ell,i})^{\zeta^{\ell,i}_{10}}p_{00}(d_{\ell,i})^{\zeta^{\ell,i}_{00}}.
\end{align*}
Taking the logarithm yields
\begin{eqnarray}\label{mle}
l_\ell(\theta_\ell)&:=&\log{\mathcal{L}_\ell(\theta_\ell|y_{\ell,1,1},\ldots,y_{\ell,1,n_{\ell,1}},\ldots,y_{\ell,k_\ell,n_{\ell,k_\ell}})}\nonumber\\
&=&\sum_{i=1}^{k_\ell} \zeta^{\ell,i}_{11}\log{p_{11}(d_{\ell,i})}+\zeta^{\ell,i}_{01}\log{p_{01}(d_{\ell,i})}+\zeta^{\ell,i}_{10}\log{p_{10}(d_{\ell,i})}+\zeta^{\ell,i}_{00}\log{p_{00}(d_{\ell,i})}
\end{eqnarray}
and the estimate $\hat \theta_\ell$ for the parameter 
$\theta_\ell $ of the  Gumbel model is obtained by maximizing this function over the parameter space ($\ell=1,2$). Note that the model estimates $\hat\beta_{\ell,1},\ \hat\gamma_{\ell,1}$ are the same as the ones obtained by maximizing the likelihood function in the univariate case \eqref{mle-univ} if $\nu=0$.

Let 
$$
\eta_{\ell}(d,\theta_{\ell}) =  
\big ( \eta_{\ell}^E (d,\theta_{\ell}), \eta_{\ell}^T (d,\theta_{\ell}) \big ) ~ = ~ \Big ( \frac{1}{1 + e^{-\beta_{\ell ,1} - \gamma_{\ell,1}  \cdot d }} ~, \frac{1}{1 + e^{-\beta_{\ell ,2} - \gamma_{\ell,2}  \cdot d }} \Big )^T
$$
denote the vector  of efficacy and toxicity curves for group $\ell =1,2$. 
We now want to ensure that claiming equivalence of both treatment groups guarantees that both, efficacy and toxicity response, do not deviate more than a certain prespecified threshold $\epsilon=(\epsilon^E,\epsilon^T)$.
Consequently the global hypotheses are given by
\begin{equation} \label{h0} 
H_0:  \max_{d\in \cal D}\left| \eta_1^E(d,\theta_1)-\eta_2^E(d,\theta_2)\right| \geq \epsilon^E \text{ or }  \max_{d\in \cal D}\left| \eta_1^T(d,\theta_1)-\eta_2^T(d,\theta_2)\right| \geq \epsilon^T \end{equation}
against the alternative
\begin{equation} \label{h1} 
H_1:  \max_{d\in \cal D}\left| \eta_1^E(d,\theta_1)-\eta_2^E(d,\theta_2)\right| < \epsilon^E \text{ and }  \max_{d\in \cal D}\left| \eta_1^T(d,\theta_1)-\eta_2^T(d,\theta_2)\right| < \epsilon^T. \end{equation}
This  problem can be solved by simultaneously testing the individual hypotheses
{\small \begin{equation} \label{hypothesesE} 
H_0^E:  \max_{d\in \cal D}\left| \eta_1^E(d,\theta_1)-\eta_2^E(d,\theta_2)\right| \geq \epsilon^E \text{ vs. } H_1^E:\max_{d\in \cal D}\left| \eta_1^E(d,\theta_1)-\eta_2^E(d,\theta_2)\right| < \epsilon^E \end{equation}}
and
{\small  \begin{equation} \label{hypothesesT} 
H_0^T:  \max_{d\in \cal D}\left| \eta_1^T(d,\theta_1)-\eta_2^T(d,\theta_2)\right| \geq \epsilon^T \text{ vs. } H_1^T:\max_{d\in \cal D}\left| \eta_1^T(d,\theta_1)-\eta_2^T(d,\theta_2)\right| < \epsilon^T.
\end{equation}}
As the global null in \eqref{h0} is the union of $H_0^E$ and $H_0^T$ we can apply the Intersection-Union-Principle (see  \cite{berger1982}). 
Only if both individual null hypotheses in \eqref{hypothesesE} and \eqref{hypothesesT} can be rejected, the global null in \eqref{h0} is rejected and equivalence of the two responses can be claimed. 
Each of the two individual tests for \eqref{hypothesesE} and \eqref{hypothesesT} is performed by extending the parametric bootstrap approach in Algorithm  \ref{alg1}, as described in the following algorithm. 

\begin{algorithm}
\label{alg2} {\rm (parametric bootstrap for testing equivalence for bivariate binary outcomes)
	\begin{itemize}
		\item[(1)]
		Calculate the MLE $\hat \theta_\ell=(\hat\beta_{\ell,1},\hat\gamma_{\ell,1},\hat\beta_{\ell,2},\hat\gamma_{\ell,2},\hat\nu_{\ell})$, $\ell=1,2$, by maximizing the log-likelihood given in \eqref{mle} for each group.
		The test statistics are obtained by
		$$\hat d^E=d^E(\hat\theta_1,\hat\theta_2)=\max_{d\in \cal D}\big| \eta_1^E(d,\hat\theta_1)-\eta_2^E(d,\hat\theta_2)\big|$$
		and
		$$\hat d^T=d^T(\hat\theta_1,\hat\theta_2)=\max_{d\in \cal D}\big| \eta_1^T(d,\hat\theta_1)-\eta_2^T(d,\hat\theta_2)\big|$$
		\item[(2)]  For each individual test for \eqref{hypothesesE} and \eqref{hypothesesT} we perform a constrained optimization as described in Algorithm \ref{alg1}, yielding estimates $\hat{\hat{\theta}}_{\ell}$, $\ell=1,2$.  Note that this procedure is done separately for each individual test because the  constraints and hence the generation of the bootstrap data differ. 
Thus we generate bootstrap data for each individual test separately and obtain  replicates $\hat d^{E*}_{\infty,1}, \dots, \hat d^{E*}_{\infty,n_{boot}}$  for the comparison of the efficacy 
curves 
and $\hat d^{T*}_{\infty,1}, \dots, \hat d^{T*}_{\infty,n_{boot}}$ for the comparison of the  toxicity curves.    Let $\hat d^{E*(1)} \le \ldots  \le \hat d^{E*(n_{boot})}$ and $\hat d^{T*(1)} \le \ldots  \le \hat d^{T*(n_{boot})}$ denote the corresponding order statistics and let $\hat d^{E*(\lfloor n_{boot} \alpha \rfloor )}$  and  $\hat d^{T*(\lfloor n_{boot}\alpha \rfloor )}$ denote the corresponding empirical level $\alpha$ quantiles.
	\item[(3)] Reject the global null hypothesis \eqref{h0}  if
		\begin{equation} \label{testInf2}
		\hat d^E < \hat d^{E*(\lfloor n_{boot} \alpha \rfloor )}~~ \text{ and } ~~
		\hat d^T < \hat d^{T*(\lfloor n_{boot}\alpha \rfloor )}. 
		\end{equation}
	\end{itemize}}
\end{algorithm}
		Note that according to the Intersection-Union-Principle we use the $\alpha$-quantile and there is no need of adjusting the level of the two individual tests. 
The technical difficulty of the implementation of this algorithm consists in generating bivariate correlated binary data in Step (2), which is explained in more detail in the following section.

\subsection{Generation of bivariate correlated binary data}\label{gen_biv_bin_data}

The bootstrap test described in Algorithm \ref{alg2} requires the simulation of bivariate binary data. Due to the dependency of the outcomes this is a technical difficulty investigated by several authors (see for example \cite{emrich1991,lunn1998,leisch1998} among many others). 
We used the algorithm developed by \cite{emrich1991}, as implemented with the function generate.binary in the R package MultiOrd (see \cite{amatya2015}). For this purpose, we use expression \eqref{corr_gumb} for the correlation and the marginal distributions in \eqref{ecurv} to generate the data at each dose level as long as the correlation does not exceed the boundaries specified by the model parameter $\theta$ given by
\begin{equation}
\max{\left( -\sqrt{\tfrac{p_1(d)p_2(d)}{(1-p_1(d))(1-p_2(d))}},-\sqrt{\tfrac{(1-p_1(d))(1-p_2(d))}{p_1(d)p_2(d)}}\right)} \leq corr(Y^E,Y^T|\ d)
\end{equation}
and
\begin{equation}\label{restr}
corr(Y^E,Y^T|\ d)\leq min{\left( \sqrt{\tfrac{p_1(d)(1-)p_2(d))}{(1-p_1(d))p_2(d)}},\sqrt{\tfrac{(1-p_1(d))p_2(d)}{p_1(d)(1-p_2(d))}}\right)}.
\end{equation}
Here, $p_1(d)=\eta^E(d,\theta_1)$ and $p_2(d)=\eta^T(d,\theta_2)$ denote the marginal probabilities of efficacy and toxicity, respectively.
These restrictions have to be fulfilled at each dose in order to guarantee that a joint distribution of $Y^E$ and $Y^T$ can exist. 
We impose these inequality constraints in the  optimization step in addition to the constraint described in \eqref{constr} such that the estimates $\hat{\hat{\theta}}_1$ and $\hat{\hat{\theta}}_2$ generate a distribution and bootstrap data can be obtained.

\section{Finite sample properties}\label{sec:simul}
\def\theequation{4.\arabic{equation}}
\setcounter{equation}{0}

We now investigate the finite sample properties of the two tests based on Algorithms \ref{alg1} and \ref{alg2}. Following \cite{murtaugh1990} we consider the (log-transformed) dose range $\mathcal{D}=[ -3,3 ] $ and $7$ dose levels $-3,-2,\ldots, 2,3$. We assume $7$, $14$, $21$, $28$ and $50$ patients per dose level and group, that is $n_{\ell,i}=7,\ 14,\ 21,\ 28$ and $50$ for $\ell=1,2$, $i=1,\ldots 7$, resulting in $n_\ell=49,98,147,196$ and $350$, $\ell=1,2$. The significance level is $\alpha=0.05$ throughout. Following \cite{nam1997} and \cite{chen2000} we assume three different equivalence thresholds, $0.1$, $0.15$ and $0.2$.
 All simulations are performed using $1000$ simulation runs and $n_{boot}=400$ bootstrap replications. The binary data are generated as described in Section \ref{gen_biv_bin_data}. We set $\nu=0$ for the univariate case.

\subsection{Univariate efficacy outcomes}\label{sec:binary-sim}

\begin{figure}[h]
	\centering
\includegraphics[width=0.7\textwidth]{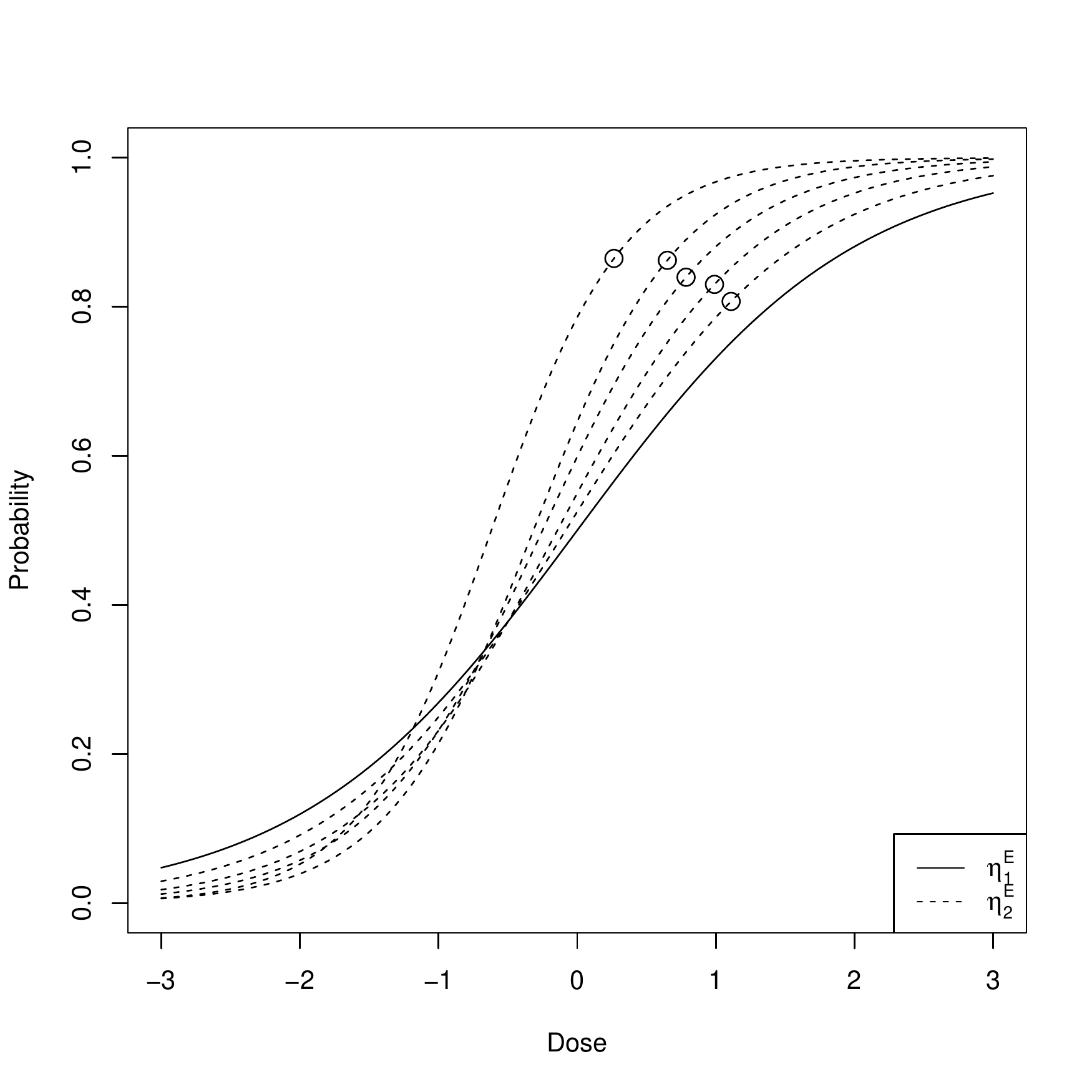} 
	\caption{\small \it The reference efficacy curve $\eta^E_1(d,\beta_{1,1},\gamma_{1,1})= \frac{1}{1+e^{-d}}$ (solid line) and the curves $\eta^E_2(d,\beta_{2,1},\gamma_{2,1})= \frac{1}{1+e^{-\beta_{2,1}-\gamma_{2,1}d}}$ (dashed lines) for different choices of the parameters $(\beta_{2,1},\gamma_{2,1})$, as defined in \eqref{scen1}. The scenarios correspond to a maximum absolute deviation $d^E= 0,\ 0.05,\ 0.1,\ 0.15,\ 0.2,\ 0.3$ attained at the open dots (from right to left).}
	\label{fig1a}
\end{figure}

We consider a logistic regression in \eqref{log_reg}. The reference model is  specified by $(\beta_{1,1},\gamma_{1,1})=(0,1)$ yielding
$
\eta^E_1(d,\beta_{1,1},\gamma_{1,1})= \frac{1}{1+e^{-d}}.
$
We choose the parameters $(\beta_{2,1},\gamma_{2,1})$ of the second model as $$\eta^E_2(d,\beta_{2,1},\gamma_{2,1})= \frac{1}{1+e^{-\beta_{2,1}-\gamma_{2,1}d}},$$ such that the maximum deviations $d^E$ between the two efficacy curves $\eta_1^E$ and $\eta_2^E$ are $ 0,\ 0.05,\ 0.1,\ 0.15,\ 0.2$ and $0.3$, attained at the doses $1.11$, $0.99$, $0.78$, $0.65$ and $0.26$, respectively. This leads to the configurations
\begin{align}\label{scen1}
(\beta_{2,1},\gamma_{2,1})= (0,1),\ (\beta_{2,1},\gamma_{2,1})= (0.1,1.2),\ (\beta_{2,1},\gamma_{2,1})= (0.2,1.4),\nonumber\\ (\beta_{2,1},\gamma_{2,1})= (0.4,1.6),\ (\beta_{2,1},\gamma_{2,1})= (0.6,1.9) \text{ and }(\beta_{2,1},\gamma_{2,1})= (1.3,2.1).
\end{align}
Note that for $\eta^E_1=\eta^E_2$ the difference between the curves is zero at all doses. 
Figure \ref{fig1a} displays the reference curve $\eta^E_1$ and the curve $\eta^E_2$ determined by the parameters described in \eqref{scen1}.

 Table \ref{tab1} displays the simulated  type I error rates of the bootstrap test \eqref{testInf} for the equivalence of  efficacy responses with margins $\epsilon^E=0.1$, $0.15$, $0.2$. The numbers in bold face indicate the scenarios where simulations have been run on the margin of the null, that is $d^E=\epsilon^E=0.1,\ 0.15$ and $0.2$. Note that the configuration $d^E=0.15$ and $\epsilon^E=0.2$ falls under the alternative (as $d^E<\epsilon^E$) and is  therefore omitted from Table \ref{tab1}.
 We conclude that the test controls its level in all cases under consideration. The approximation of the level is very precise at the margin of the null hypothesis (that is, $d^E=\epsilon^E$) and this accuracy increases with increasing sample sizes. 
 Moreover, in the interior of the null hypothesis (that is $d^E\geq\epsilon^E$) the number of rejections is close to zero in all scenarios, indicating that the type I error rate is well below $\alpha$ in these cases.
 
 Table \ref{tab2} displays the power of the test \eqref{testInf}. We conclude that for sufficiently large sample sizes the procedure has reasonable power. For instance, for $n_{\ell,i}=28,\ i=1,\ldots 7,\ \ell=1,2,$ the maximum power attained at  $d^E=0$ is $0.785$ for an equivalence threshold of $\epsilon^E=0.2$. For larger sample sizes of $50$ patients per dose level, the test achieves more than $80\%$ power, namely $0.803$ for $\epsilon^E=0.15$ and $0.976$ for $\epsilon^E=0.2$. 
 In general we observe that the power increases with increasing sample sizes.
 Note that the case $d^E=\epsilon^E=0.1$ falls under the null hypothesis and results are therefore shown in Table \ref{tab1}.

\renewcommand{\arraystretch}{0.7}
\begin{table}[ht]
	{ \small
		\centering
\begin{tabular}{||c|c|c|c|c|c|c||}
	\hline
	$n_{\ell,i}$ & $(\beta_{2,1},\gamma_{2,1})$ & $d^E$ & $\epsilon^E=0.1$ & $\epsilon^E=0.15$ & $\epsilon^E=0.2$	\\ \hline
\multirow{4}{*}{$7$} & (1.3,2.1)& $0.3$&  0.005  & 0.014 &0.018    \\	
			& (0.6,1.9) & $0.2$&  0.020 & 0.022 &\textbf{0.055 }  \\
			&  (0.4,1.6)& $0.15$&  0.036  & \textbf{0.037 } & - \\
			&  (0.2,1.1)& $0.1$&  \textbf{0.060}  & - & - \\\hline
			\multirow{4}{*}{$14$} & (1.3,2.1) & $0.3$&  0.002  &   0.001 & 0.003\\	
			& (0.6,1.9)& $0.2$& 0.005  & 0.014 &\textbf{0.055}\\
			&  (0.4,1.6)& $0.15$&  0.020  &   \textbf{0.038 }&- \\
		    &  (0.2,1.1)& $0.1$&  \textbf{0.042}  & - & - \\\hline
			\multirow{4}{*}{$21$} &(1.3,2.1) & $0.3$& 0.000  & 0.000  & 0.002  \\	
			&  (0.6,1.9)& $0.2$&  0.004 & 0.007 & \textbf{0.052} \\
			& (0.4,1.6)& $0.15$& 0.010  &   \textbf{0.042}  & - \\
			&  (0.2,1.1)& $0.1$&  \textbf{0.036}  & - & - \\\hline
			\multirow{4}{*}{$28$} & (1.3,2.1)& $0.3$& 0.000  &0.000 & 0.001  \\	
			& (0.6,1.9)& $0.2$& 0.000  & 0.012 & \textbf{0.062} \\
			& (0.4,1.6)& $0.15$& 0.008 &  \textbf{0.040} & - \\
			&  (0.2,1.1)& $0.1$&  \textbf{0.036}  & - & - \\\hline
			\multirow{4}{*}{$50$} & (1.3,2.1) & $0.3$& 0.000 & 0.000   & 0.000 \\	
			&  (0.6,1.9)& $0.2$& 0.002  & 0.011 &\textbf{0.057}  \\
			&  (0.4,1.6)& $0.15$& 0.006   &  \textbf{0.052} & - \\
			&  (0.2,1.1)& $0.1$&  \textbf{0.034}  & - & - \\\hline\hline
		\end{tabular}
		\caption{\label{tab1} \it Simulated type I error rates  of the bootstrap test \eqref{testInf} for the equivalence of  efficacy responses. 
		Bold numbers indicate simulations at the margin of the null hypothesis.}
	}
\end{table}

\begin{table}[ht]
	{\small
		\centering
		\begin{tabular}{||c|c|c|c|c|c|c||}
			\hline
		 $n_{\ell,i}$ & $(\beta_{2,1},\gamma_{2,1})$ & $d^E$ & $\epsilon^E=0.1$ & $\epsilon^E=0.15$ & $\epsilon^E=0.2$	\\ \hline
			\multirow{3}{*}{$7$} & (0.2,1.4) & $0.1$&  -  & 0.082 & 0.090  \\	
			& (0.1,1.2) & $0.05$& 0.058  & 0.076 & 0.145 \\
			&  (0,1)& $0$& 0.076  & 0.171 & 0.232 \\\hline
			\multirow{3}{*}{$14$} & (0.2,1.4) & $0.1$&  -  & 0.137  &0.247 \\	
			& (0.1,1.2) & $0.05$&  0.075 &  0.142&0.391 \\
			&  (0,1)& $0$&   0.101 &   0.226 & 0.418\\\hline
			\multirow{3}{*}{$21$} &(0.2,1.4)  & $0.1$&  - & 0.166 & 0.434 \\	
			&  (0.1,1.2)& $0.05$&  0.090 & 0.344 &0.547 \\
			& (0,1)& $0$& 0.134  &   0.356 & 0.603 \\\hline
			\multirow{3}{*}{$28$} & (0.2,1.4) & $0.1$& -  & 0.203 & 0.474 \\	
			&  (0.1,1.2)& $0.05$&  0.103 & 0.367& 0.690\\
			& (0,1)& $0$& 0.179 &  0.470  & 0.785 \\\hline 	
			\multirow{3}{*}{$50$} & (0.2,1.4) & $0.1$& - &  0.303  & 0.729\\	
			&  (0.1,1.2)& $0.05$&  0.184 & 0.640 & 0.905 \\
			&  (0,1)& $0$&  0.363  &  0.803 & 0.976 \\
			\hline 			 		  
			\hline
		\end{tabular}
		\caption{\label{tab2} \it Simulated power of the bootstrap test \eqref{testInf} for the equivalence of efficacy responses. 
		}
	}
\end{table}
\renewcommand{\arraystretch}{1}

\subsection{Bivariate efficacy-toxicity outcomes}\label{sec:gumbel-sim}

\begin{figure}[h]
	\centering
    \subfigure{\includegraphics[width=0.49\textwidth]{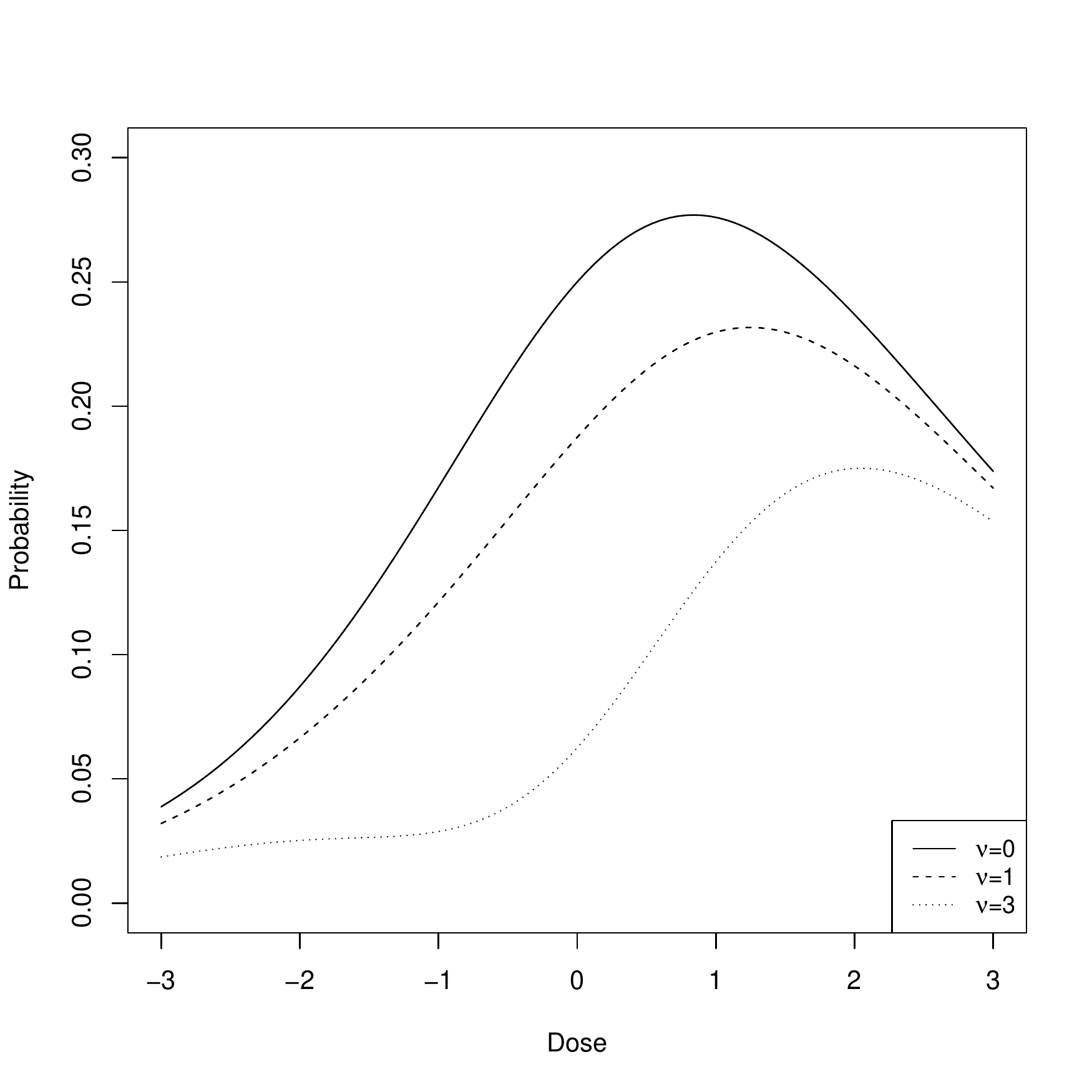}} 
	\subfigure{\includegraphics[width=0.49\textwidth]{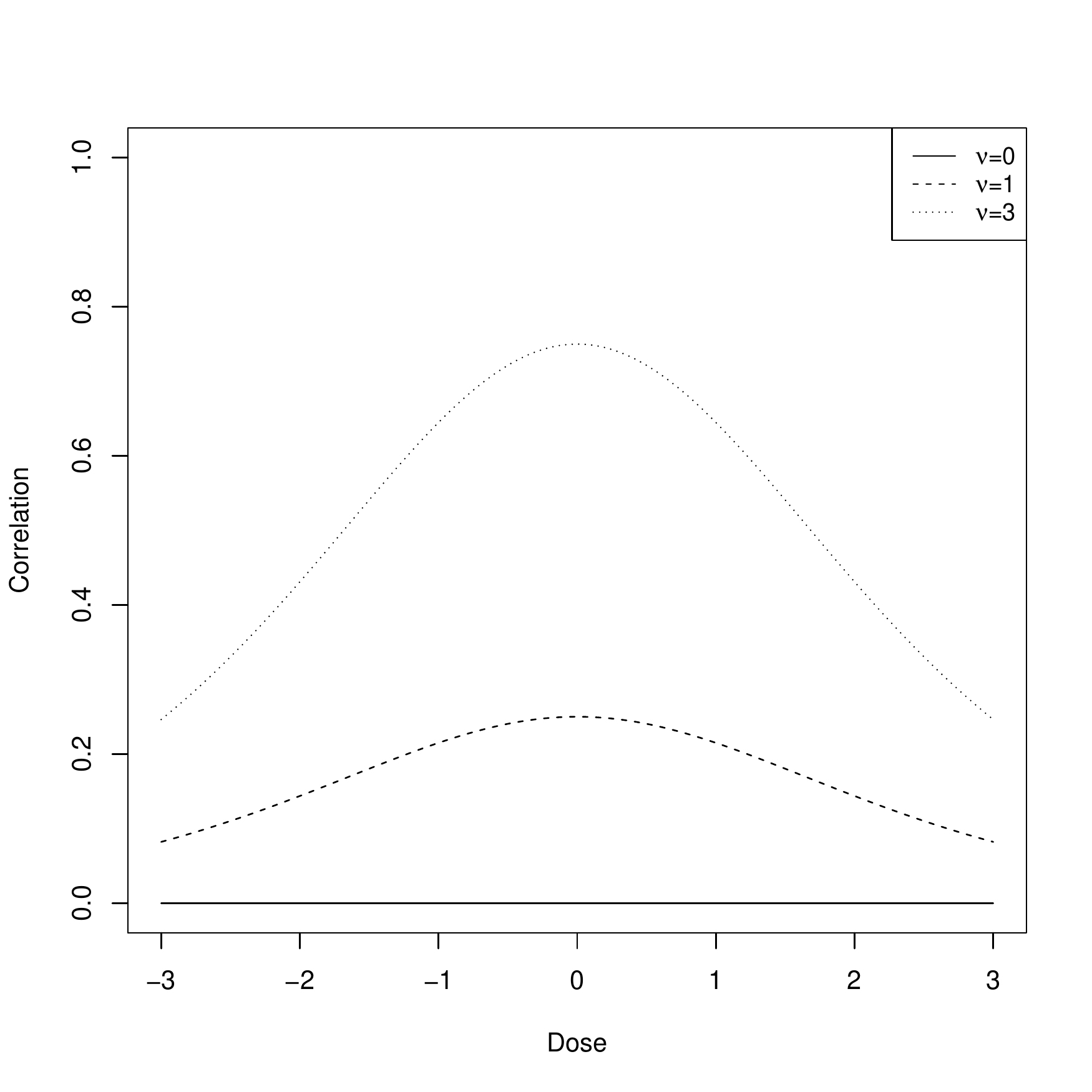}} 
	\caption{\small \it Left panel: Probability $\mathbb{P}(Y^E=1, Y^T=0 )=p_{10}(d)$ in dependence of the dose for the reference model \eqref{ref_gumb} for different choices of the correlation parameter $\nu$. Right panel: Correlation of efficacy and toxicity response for different choices of $\nu$ in dependence of the dose.}
	\label{fig1b}
\end{figure}

\begin{figure}[ht]
	\subfigure{\includegraphics[width=0.32\textwidth]{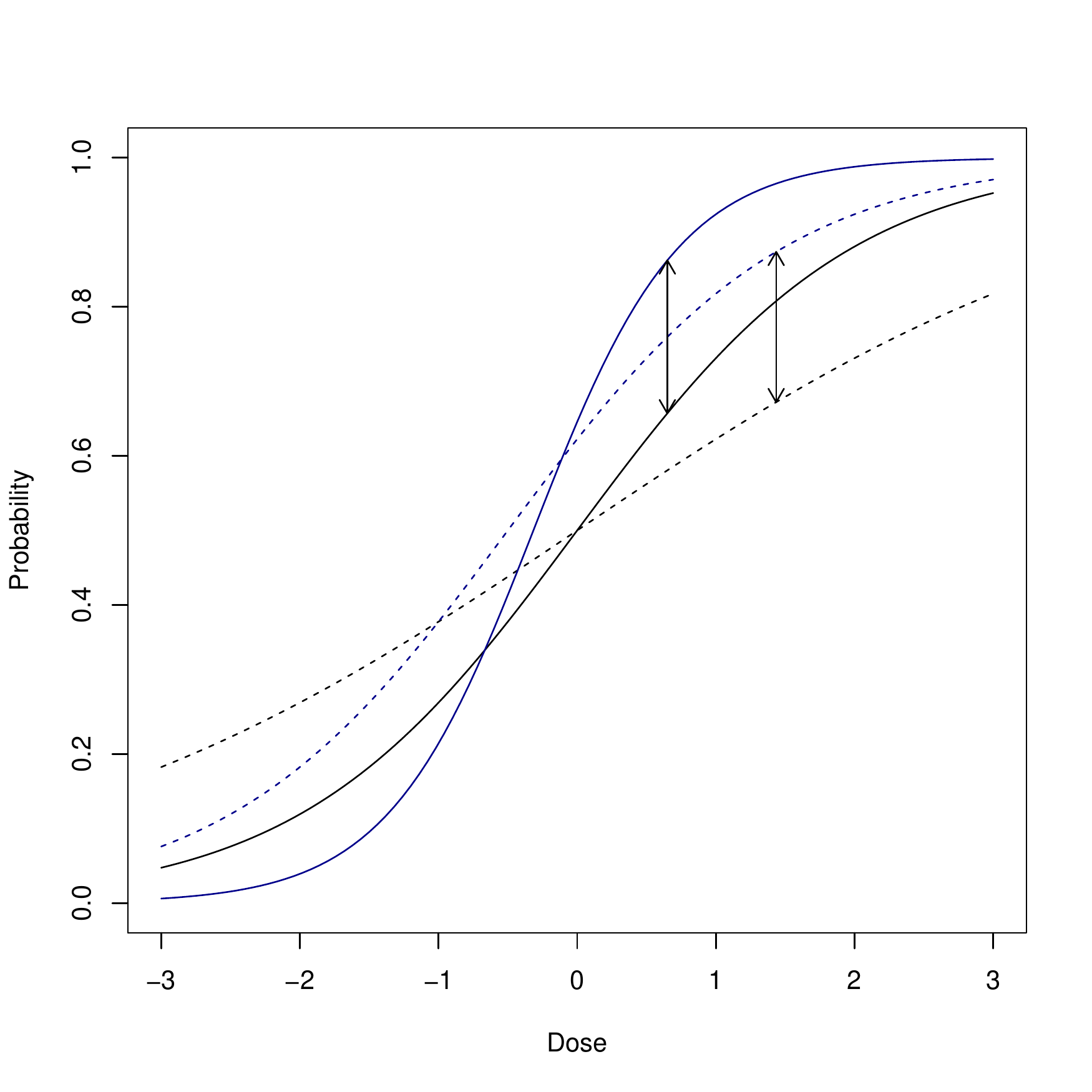}} 
	\subfigure{\includegraphics[width=0.32\textwidth]{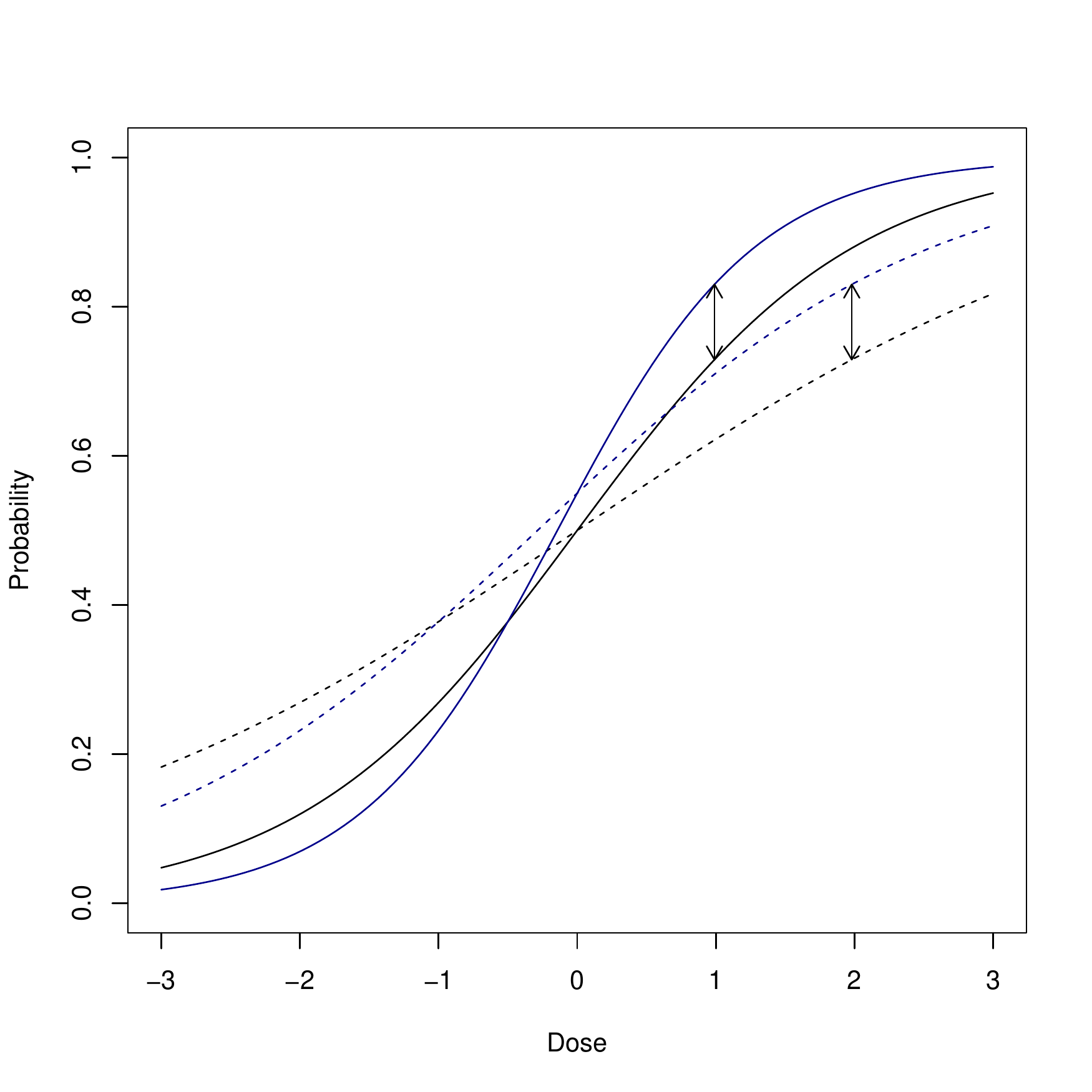}} 
	\subfigure{\includegraphics[width=0.32\textwidth]{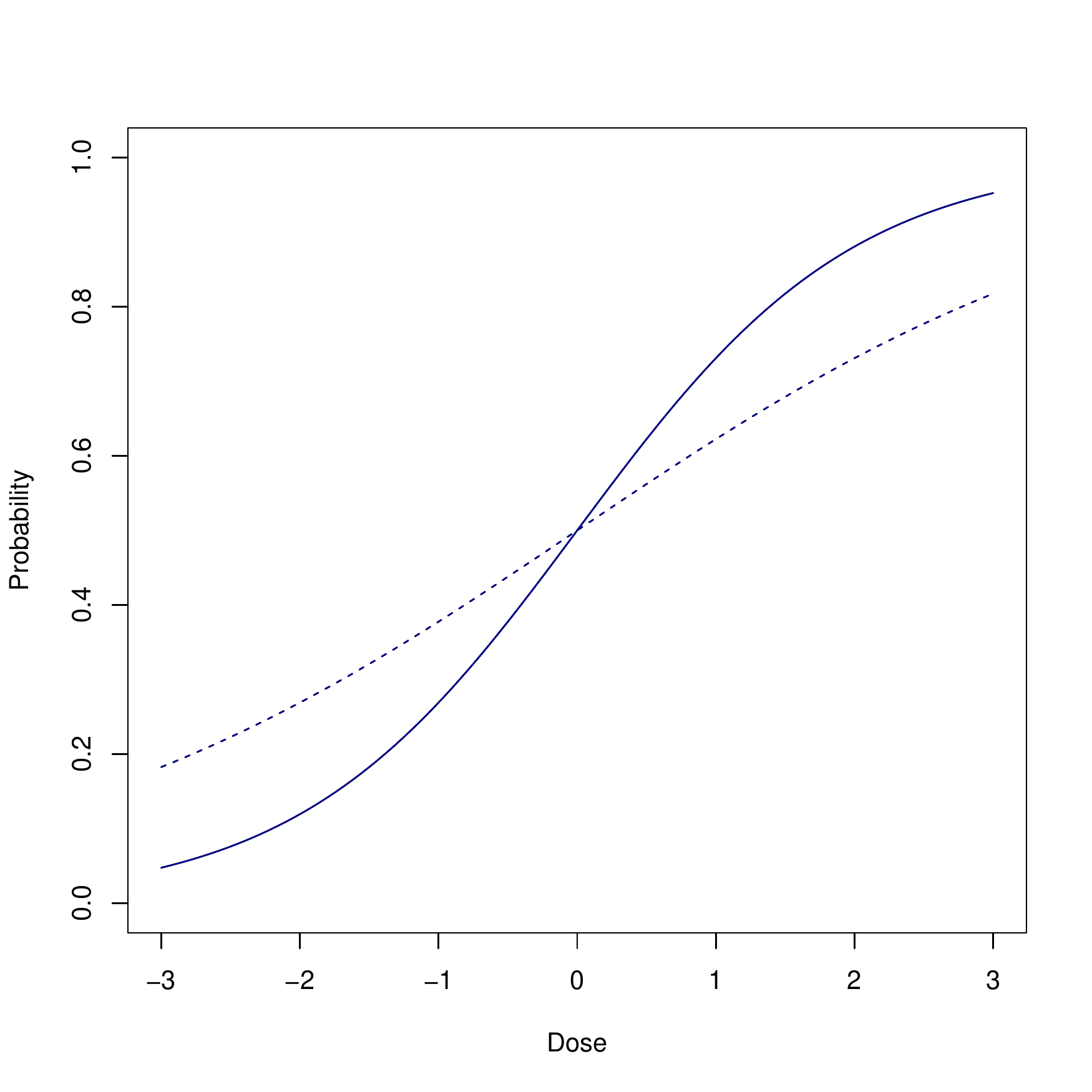}} 
	\caption{\small \it Efficacy curves (solid lines) and toxicity curves (dashed lines) derived in \eqref{ecurv}. The black lines correspond to the reference model, the blue lines to the second model, specified by $\theta_2$. The scenarios shown correspond to a maximum absolute deviation (indicated by the arrows) of $d^E=d^T=0.2,\ 0.1$ and $0$ (from left to right).}
	\label{fig2}
\end{figure}

We now consider bivariate efficacy-toxicity outcomes using a Gumbel model for both treatment groups as defined in Section \ref{sec:gumbel}. The reference model is defined by the parameter 
\be\theta_1=(\beta_{1,1},\gamma_{1,1},\beta_{1,2},\gamma_{1,2},\nu_1)=(0,1,0,0.5,\nu_1)\label{ref_gumb}\ee
and
we assume two different levels of dependence representing a moderate ($\nu_1=1$) and a rather strong dependence ($\nu_1=3$) between the efficacy and toxicity outcomes. According to \eqref{corr_gumb}, the correlation of $Y_1^E$ and $Y_1^T$ at dose $d \in \mathcal{D}$ is given by 
\be
corr(Y_1^E, Y_1^T|\ d)=\frac{\nu_1}{(e^{d/2}+e^{-d/2})(e^{d/4}+e^{-d/4})},
\ee
which ranges from $0.08$ to $0.25$ for $\nu_1=1$ and $0.25$ to $0.75$ for $\nu_1=3$. Note that the highest correlation is always attained at the dose level $0$. The left panel of Figure  \ref{fig1b} displays the probability of efficacy without toxicity response, that is $\mathbb{P}(Y^E=1, Y^T=0|\ d)=p_{10}(d)$. The right panel displays the correlation for different choices of $\nu$ in dependence of the dose.
In order to investigate different situations under the null and the alternative, we vary the parameters of the second model resulting in seven scenarios for each choice of $\nu_1$; see Table \ref{tab_par}.  We assume the same correlations as for the reference model, that is $\nu_2=\nu_1$. As an illustration, we show the efficacy and toxicity curves for three scenarios and $\nu_1=1$ in Figure \ref{fig2}.

\renewcommand{\arraystretch}{0.9}
\begin{table}[ht]
	{\small
		\centering
		\begin{tabular}{||c|c|c|c||}
			\hline
		& $\theta_1$ & $\theta_2$ & $d=(d^E,d^T)$ 	\\ \cline{1-4}
			\multirow{3}{*}{Alternative} & $(0,1,0,0.5,\nu_2)$&$(0,1,0,0.5,\nu_2)$ &  $(0,0)$  \\	
			&$(0,1,0,0.5,\nu_2)$  &$(0.1,1.2,0.1,0.6,\nu_2)$ & $(0.05,0.05)$ \\
			& $(0,1,0,0.5,\nu_2)$ & $(0.2,1.4,0.2,0.7,\nu_2)$ & $(0.1,0.1)$ \\ \hline
				\multirow{4}{*}{Null hypothesis} & $(0,1,0,0.5,\nu_2)$ &$(0.4,1.6,0.4,0.8,\nu_2)$ &  $(0.15,0.15)$  \\	
			& $(0,1,0,0.5,\nu_2)$ & $(0,1,0.4,0.8,\nu_2)$& $(0,0.15)$ \\
			& $(0,1,0,0.5,\nu_2)$ & $(0.6,1.9,0.5,1,\nu_2)$ & $(0.2,0.2)$  \\
			& $(0,1,0,0.5,\nu_2)$ &$(0,1,0.5,1,\nu_2)$ & $(0,0.2)$  \\\hline\hline
				\end{tabular}
		\caption{\label{tab_par} \it Different  scenarios corresponding to the null hypothesis \eqref{h0} and the alternative \eqref{h1}.  }
	}
\end{table}

For the Type I error rate simulations we counted the number of individual and simultaneous rejections of both null hypotheses in \eqref{hypothesesE} and \eqref{hypothesesT}, allowing us to reject the global null hypothesis in \eqref{h0}. All simulation results are displayed in Tables \ref{tab3} and \ref{tab4}, where the numbers in brackets correspond to the proportion of rejections for the individual tests on efficacy and toxicity. For the sake of brevity we assume only two different thresholds $\epsilon=(\epsilon^E,\epsilon^T)=(0.15,0.15)$ and $(0.2,0.2)$, thus allowing for a deviation of $15\%$ and $20\%$, respectively, for efficacy and toxicity in order to claim equivalence.
In general, we observe that the global bootstrap test according to Algorithm \ref{alg2} is rather conservative as the Type I error rates are very small. 
For example, for $n_{\ell,i}=21$, $\nu_1=\nu_2=1$ and $\epsilon=(0.2,0.2)$ the individual proportions of rejection are $0.041$ for efficacy and $0.050$ for toxicity, whereas the Type I error rate for the global test is $0.005$, which is well below the nominal level.
This is a common feature of the Intersection-Union-Principle for the problem of testing bioequivalence in multivariate responses (see, for example \cite{berger1996}). 

In general, we conclude that the individual tests on efficacy and toxicity yield rejection probabilities that are very close to 0.05 when simulating on the margin of the global null hypothesis (that is $d=\epsilon$) and hence the global Type I error rates are well below $\alpha$ in these cases. However, there are some scenarios where the Type I error rate is too large when $\nu_1=\nu_2=3$. For instance, we observe a proportion of rejections of the global null hypothesis of $0.129$ for $n_{\ell,i}=28$, $\epsilon=(0.2,0.2)$ and $d=(0,0.2)$.
Note that the values of $\nu_1$ and $\nu_2$ do not influence the curves obtained by the marginal densities in \eqref{marg}  and hence do not directly impact the proportions of rejections obtained for the individual tests. However, the choice of $\nu_\ell$ affects the estimation of the  parameter $\theta_\ell$ of the Gumbel model, which explains the different results for the individual tests for $\nu_\ell=1$ and $\nu_\ell=3$, $\ell=1,2$, resulting in higher Type I error rates for the global test in settings with $\nu_\ell=3$. For example, the choice $d=(0.2,0)$ and $\epsilon=(0.2,0.2)$ corresponds to the global null hypothesis \eqref{h0}, as   $d^E=0.2=\epsilon^E$. However, due to the fact that we are far under the alternative for toxicity ($d^T=0$) and due to the high correlation ($\nu_\ell=3$) we observe a Type I error rate inflation for the individual test on efficacy and consequently for the global test as well, for all sample sizes. There are two reasons causing this effect: on the one hand the high correlation results in difficulties to estimate the curves properly, even for large sample sizes. On the other hand, the asymptotic distribution of the maximum absolute deviation of the two curves is different and more complex in case of $d^T=0$, which also affects the test results; see \cite{detmolvolbre2015} for further numerical and theoretical details on this issue.

A similar argument also holds for the power results shown in Table \ref{tab4}. It turns out that the global test achieves reasonable power for sufficiently large sample sizes. For example a maximum power (always attained at $d=(0,0)$) of $0.817$ is achieved for the global test for a choice of $n_{\ell,i}=28$, $\nu_1=\nu_2=3$ and $\epsilon=(0.2,0.2)$. For a lower threshold, that is  $\epsilon=(0.15,0.15)$, the maximum power is smaller, but still increasing with growing sample sizes, reaching for instance $0.830$ for $n_{\ell,i}=50$ and $\nu_1=\nu_2=3$. 
\renewcommand{\arraystretch}{0.8}
\begin{table}[h!]
	{
		\centering\footnotesize
		\begin{tabular}{||c|c|c|c|c|c||}
			\hline
	$\epsilon=(\epsilon^E,\epsilon^T)$ &	 $n_{\ell,i}$ & $\theta_2$ & $d=(d^E,d^T)$  & $\nu_\ell=1$ & $\nu_\ell=3$	\\ \cline{1-6}
\multirow{9}{*}{$(0.15,0.15)$}  &\multirow{2}{*}{$7$} & $ (0.4,1.6,0.4,0.8,\nu_2)$ & $(0.15,0.15)$  & 0.004 (0.051/0.050)  &  0.021 (0.078/0.063)  \\	
&& $ (0,1,0.4,0.8,\nu_2)$ & $(0,0.15)$  & 0.009 (0.142/0.056) & 0.029 (0.180/0.087) \\\cline{2-6}
&\multirow{2}{*}{$14$} & $ (0.4,1.6,0.4,0.8,\nu_2)$ & $(0.15,0.15)$  & 0.005 (0.052/0.056) & 0.006 (0.041/0.044) \\	
&& $ (0,1,0.4,0.8,\nu_2)$ & $(0,0.15)$  & 0.007 (0.212/0.051) & 0.025 (0.256/0.071) \\\cline{2-6}
&\multirow{2}{*}{$21$} & $ (0.4,1.6,0.4,0.8,\nu_2)$ & $(0.15,0.15)$  & 0.005 (0.032/0.055) & 0.011 (0.042/0.050)   \\	
&& $ (0,1,0.4,0.8,\nu_2)$ & $(0,0.15)$  & 0.029 (0.364/0.056) &  0.049 (0.395/0.091)  \\\cline{2-6}
&\multirow{2}{*}{$28$} & $ (0.4,1.6,0.4,0.8,\nu_2)$ & $(0.15,0.15)$  & 0.004 (0.036/0.044) & 0.014 (0.055/0.065)   \\	
&& $ (0,1,0.4,0.8,\nu_2)$ & $(0,0.15)$  & 0.017 (0.648/0.044) & 0.064 (0.610/0.098)  \\\cline{2-6}		 	
&\multirow{2}{*}{$50$} & $ (0.4,1.6,0.4,0.8,\nu_2)$ & $(0.15,0.15)$ & 0.004 (0.050/0.047) &  0.035 (0.096/0.071)  \\	
&& $ (0,1,0.4,0.8,\nu_2)$ & $(0,0.15)$   & 0.053 (0.831/0.062) & 0.112 (0.866/0.128)  \\
\hline 			 		  
\hline
\multirow{10}{*}{$(0.2,0.2)$} & \multirow{2}{*}{$7$} & $(0.6,1.9,0.5,1,\nu_2)$ & $(0.2,0.2)$  & 0.002 (0.061/0.038)  & 0.021 (0.062/0.064)  \\	
&& $ (0,1,0.5,1,\nu_2)$ & $(0,0.2)$  & 0.018 (0.209/0.060) & 0.029 (0.253/0.077) \\\cline{2-6}
&\multirow{2}{*}{$14$} & $(0.6,1.9,0.5,1,\nu_2)$ & $(0.2,0.2)$  & 0.007 (0.033/0.042)  &  0.009 (0.042/0.046) \\	
&& $ (0,1,0.5,1,\nu_2)$ & $(0,0.2)$  & 0.015 (0.417/0.038) &  0.053 (0.465/0.088)  \\\cline{2-6}
&\multirow{2}{*}{$21$} & $(0.6,1.9,0.5,1,\nu_2)$ & $(0.2,0.2)$  & 0.005 (0.041/0.050) & 0.018 (0.060/0.057)  \\	
&& $ (0,1,0.5,1,\nu_2)$ & $(0,0.2)$  & 0.025 (0.431/0.034) &  0.074 (0.612/0.093)  \\\cline{2-6}
&\multirow{2}{*}{$28$} & $(0.6,1.9,0.5,1,\nu_2)$ & $(0.2,0.2)$  &0.008 (0.068/0.072) & 0.021 (0.070/0.050)  \\	
&& $ (0,1,0.5,1,\nu_2)$ & $(0,0.2)$  & 0.044 (0.796/0.055) & 0.129 (0.817/0.144)   \\\cline{2-6}		 	
&\multirow{2}{*}{$50$} & $(0.6,1.9,0.5,1,\nu_2)$ & $(0.2,0.2)$ & 0.009 (0.076/0.067) &  0.043 (0.103/0.083)  \\	
&& $ (0,1,0.5,1,\nu_2)$ & $(0,0.2)$   & 0.053 (0.968/0.059) & 0.223 (0.972/0.267)   \\\hline 			 		  
\hline
		\end{tabular}
		\caption{\label{tab3}\small \it Simulated Type I error rates of the global bootstrap test \eqref{testInf2}  for different choices of $\nu_\ell,\ \ell=1,2$. The numbers in brackets show the proportion of rejections for the individual tests according to  the hypotheses  \eqref{hypothesesE} and \eqref{hypothesesT}.  }
	}
\end{table}

\begin{table}[h!]
	{
\centering\footnotesize
\begin{tabular}{||c|c|c|c|c|c||}
		\hline
			$\epsilon=(\epsilon^E,\epsilon^T)$ & $n_{\ell,i}$ & $\theta_2$ & $d=(d^E,d^T)$  & $\nu_\ell=1$ & $\nu_\ell=3$	\\ \cline{1-6}
		\multirow{14}{*}{$(0.15,0.15)$}  &\multirow{3}{*}{$7$} & $ (0.2,1.4,0.2,0.7,\nu_2)$ & $(0.1,0.1)$  & 0.005 (0.076/0.088) & 0.020 (0.089/0.092) \\	
		&& $ (0.1,1.2,0.1,0.6,\nu_2)$ & (0.05,0.05)   & 0.010 (0.117/0.109)  & 0.047 (0.168/0.142) \\	
		&& $(0,1,0,0.5,\nu_2)$ & $(0,0)$ & 0.015 (0.120/0.133) & 0.061 (0.179/0.148) \\\cline{2-6} 
		&\multirow{3}{*}{$14$} & $(0.2,1.4,0.2,0.7,\nu_2)$ & $(0.1,0.1)$  & 0.008 (0.144/0.119) & 0.042 (0.137/0.116)\\	
		&& $ (0.1,1.2,0.1,0.6,\nu_2)$ & $(0.05,0.05)$& 0.027 (0.156/0.159) & 0.097 (0.236/0.204)    \\	
		&& $(0,1,0,0.5,\nu_2)$ & $(0,0)$  & 0.040 (0.234/0.182)&0.152 (0.296/0.263) \\\cline{2-6}
		&\multirow{3}{*}{$21$} & $(0.2,1.4,0.2,0.7,\nu_2)$ & $(0.1,0.1)$  & 0.023 (0.153/0.157) & 0.088 (0.197/0.189)\\	
		&& $ (0.1,1.2,0.1,0.6,\nu_2)$ & $(0.05,0.05)$ & 0.067 (0.270/0.271)  & 0.207 (0.387/0.326)    \\	
		&& $(0,1,0,0.5,\nu_2)$ & $(0,0)$ & 0.126 (0.380/0.308)& 0.259 (0.462/0.363) \\\cline{2-6} 
		&\multirow{3}{*}{$28$} & $(0.2,1.4,0.2,0.7,\nu_2)$ & $(0.1,0.1)$  & 0.029 (0.204/0.161) & 0.150 (0.262/0.266) \\	
		 && $ (0.1,1.2,0.1,0.6,\nu_2)$ & $(0.05,0.05)$ &  0.113 (0.319/0.353) & 0.344 (0.536/0.473)  \\	
		&& $(0,1,0,0.5,\nu_2)$ & $(0,0)$ &0.230 (0.502/0.441) & 0.437 (0.646/0.526) \\\cline{2-6}	
		&\multirow{3}{*}{$50$} & $(0.2,1.4,0.2,0.7,\nu_2)$ & $(0.1,0.1)$ & 0.103 (0.313/0.332) & 0.281 (0.426/0.416) \\	
&& $ (0.1,1.2,0.1,0.6,\nu_2)$ & $(0.05,0.05)$& 0.401 (0.615/0.624)  & 0.650 (0.792/0.727)   \\	
		&& $(0,1,0,0.5,\nu_2)$ & $(0,0)$ & 0.678 (0.811/0.827)& 0.830 (0.943/0.856) \\\hline 	
		\hline
			\multirow{15}{*}{$(0.2,0.2)$} &\multirow{3}{*}{$7$} & $ (0.2,1.4,0.2,0.7,\nu_2)$ & $(0.1,0.1)$  & 0.010 (0.127/0.113) & 0.060 (0.176/0.169)\\	
		&& $  (0.1,1.2,0.1,0.6,\nu_2)$ & (0.05,0.05)   & 0.037 (0.199/0.133) & 0.067 (0.236/0.177) \\	
		&& $ (0,1,0,0.5,\nu_2)$ & $(0,0)$ & 0.050 (0.230/0.186) & 0.099 (0.264/0.218)\\\cline{2-6} 
		&\multirow{3}{*}{$14$} & $ (0.2,1.4,0.2,0.7,\nu_2)$ & $(0.1,0.1)$  & 0.067 (0.197/0.239) & 0.191 (0.354/0.322) \\	
		&& $ (0.1,1.2,0.1,0.6,\nu_2)$ & $(0.05,0.05)$& 0.144 (0.351/0.353)  & 0.249 (0.495/0.409)   \\	
		&& $  (0,1,0,0.5,\nu_2)$ & $(0,0)$  & 0.198 (0.442/0.407) & 0.324 (0.560/0.439) \\\cline{2-6}
		&\multirow{3}{*}{$21$} & $ (0.2,1.4,0.2,0.7,\nu_2)$ & $(0.1,0.1)$  & 0.137 (0.317/0.365) & 0.311 (0.476/0.453) \\	
		&& $   (0.1,1.2,0.1,0.6,\nu_2)$ & $(0.05,0.05)$ & 0.286 (0.532/0.535) & 0.495 (0.683/0.578)   \\	
		&& $  (0,1,0,0.5,\nu_2)$ & $(0,0)$ &0.418 (0.676/0.613) & 0.601 (0.828/0.666) \\\cline{2-6} 
		&\multirow{3}{*}{$28$} & $ (0.2,1.4,0.2,0.7,\nu_2)$ & $(0.1,0.1)$  & 0.253 (0.483/0.478) & 0.460 (0.634/0.574) \\	
		&& $ (0.1,1.2,0.1,0.6,\nu_2)$ & $(0.05,0.05)$ & 0.451 (0.637/0.706) & 0.723 (0.870/0.775)  \\	
		&& $  (0,1,0,0.5,\nu_2)$ & $(0,0)$ & 0.650 (0.798/0.791) & 0.817 (0.942/0.843)\\\cline{2-6}	
		&\multirow{3}{*}{$50$} & $ (0.2,1.4,0.2,0.7,\nu_2)$ & $(0.1,0.1)$ & 0.511 (0.702/0.700) & 0.745 (0.853/0.804) \\	
		&& $ (0.1,1.2,0.1,0.6,\nu_2)$ & $(0.05,0.05)$& 0.826 (0.906/0.910)  &  0.964 (0.999/0.966)  \\	
		&& $  (0,1,0,0.5,\nu_2)$ & $(0,0)$ & 0.961 (0.979/0.980)& 0.985 (1.000/0.987)  \\\hline 	
		\hline
	\end{tabular}
		\caption{\label{tab4}\small \it Simulated power of the global bootstrap test \eqref{testInf2}   for different choices of $\nu_\ell,\ \ell=1,2$. The numbers in brackets show the proportion of rejections for the individual tests according to  the hypotheses  \eqref{hypothesesE} and \eqref{hypothesesT}. }
	}
\end{table}
\renewcommand{\arraystretch}{1}

\section{Case study}
\label{ssec:casestudy}
\def\theequation{5.\arabic{equation}}
\setcounter{equation}{0}
To illustrate the proposed methodology, we consider an example that is inspired by a recent consulting project of one of the authors. A nonsteroidal anti-inflammatory drug is to be investigated for its ability to attenuate dental pain after the removal of two or more impacted third molar teeth. Dental pain is a common and inexpensive setting for analgesic proof of concept, recruitment being fast and the end-point being available within a few hours. It is common to measure the pain intensity on an ordinal scale at baseline and several times after the administration of a single dose. The pain intensity difference from baseline (PID), averaged over several hours after drug administration, may then be compared with a clinical relevance threshold to create a binary success variable for efficacy. In this particular setting, side effects such as nausea and sedation after dosing were anticipated, resulting in a binary toxicity variable whether the patient experienced any such adverse events. As approved analgesics with an identified dosing range and a known dose-response relationship for tolerability are available, the objective of the study at hand was to demonstrate equivalence with a marketed product for the bivariate efficacy-toxicity outcome in a proof of concept setting.

This was a randomized double-blind parallel group trial with a total of 300 patients being allocated to either placebo or one of four active doses coded as $0.05$, $0.20$, $0.50$, and $1$ (for the new treatment) and $0.10$, $0.30$, $0.60$, and $1$ (for the marketed product), resulting in $n = 30$ per group (assuming equal allocation). To maintain confidentiality, the actual doses have been scaled to lie within the $[0, 1]$ interval. Since the study has not been completed yet, we use a hypothetical data set to illustrate the proposed methodology. 

\begin{figure}[h]
	\centering
	\includegraphics[width=0.7\textwidth]{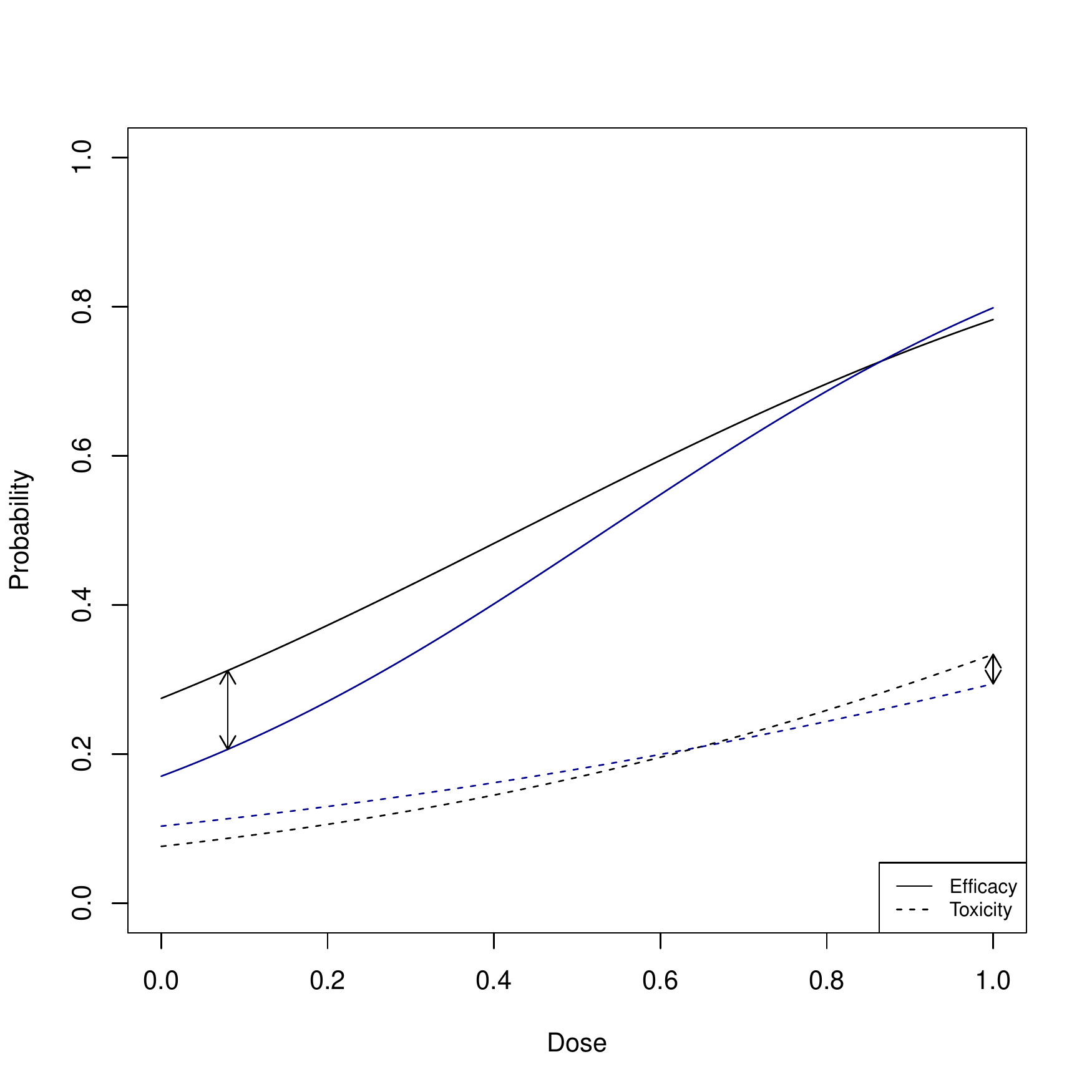}
	\caption{\small \it Efficacy and toxicity curves corresponding to the fitted Gumbel model \eqref{gumbel_case}. The black lines correspond to the marketed product, the blue lines to the new product respectively, where solid lines display the efficacy and dashed lines the toxicity response. The arrows indicate the maximum absolute distances. }
	\label{fig4}
\end{figure}

We fit two Gumbel models as defined in Section \ref{sec:gumbel} to the data, one for the marketed product ($\ell=1$) and one for the new product ($\ell=2$). The estimated model parameters are
\be\label{gumbel_case}
\hat\theta_1=(-0.971, 2.254,  -2.497, 1.806, -0.030)
,\ \hat\theta_2=(-1.585,  2.963, -2.162, 1.287, 1.003),
\ee
see Figure \ref{fig4} for the corresponding efficacy and toxicity curves.

The maximum distances are $\hat d^E=0.106$ and $\hat d^T=0.039$, attained at dose $0.08$ and the maximum dose $1$, respectively. 
We perform an equivalence test at a significance level of $\alpha=0.05$, as defined in Algorithm \ref{alg2}, for three different sets of hypotheses as we vary the equivalence thresholds $\epsilon=(\epsilon^E, \epsilon^T)$ in \eqref{hypothesesE} and \eqref{hypothesesT}.
Table \ref{tab5} displays the critical values obtained by $n_{boot}=1000$ bootstrap replications  for the different choices of $\epsilon$.

\begin{table}[h!]
		\centering\small
		\begin{tabular}{||c|c|c|c||}
			\hline
	Quantile	&	$\epsilon=(0.1,0.1)$ & $\epsilon=(0.15,0.15)$ & $\epsilon=(0.2,0.2)$  	\\ \cline{1-4}
$\hat q_{0.05}^E$ & 0.042 (0.327) & 0.073 (0.160) &0.111 (0.040)\\
$\hat q_{0.05}^T$ & 0.026 (0.094) &0.037 (0.061) & 0.054 (0.030)\\ \hline\hline
\end{tabular}
\caption{\label{tab5}\small \it Critical values of the two individual bootstrap tests on the hypotheses \eqref{hypothesesE} and \eqref{hypothesesT} for three different equivalence thresholds $\epsilon$. The numbers in brackets correspond to the p-values of the individual tests.}
\end{table}

We now test the global null hypothesis \eqref{h0} against the alternative \eqref{h1}. For $\epsilon=(0.2,0.2)$ we have  $\hat d^E=0.106< 0.111=\hat q_{0.05}^E$ and $\hat d^T=0.039< 0.054=\hat q_{0.05}^T$. According to \eqref{hypothesesE} and \eqref{hypothesesT}, we can therefore reject  \eqref{h0} at level $\alpha = 0.05$ and for $\epsilon=(0.2,0.2)$. However, we cannot reject \eqref{h0} for the other choices of $\epsilon$. For example,  $\hat d^E=0.106> 0.073=\hat q_{0.05}^E$ for  $\epsilon^E=0.1$ and we cannot reject \eqref{h0} according to \eqref{hypothesesE}. 

We obtain the same conclusions based on the observed p-values reported in brackets in Table \ref{tab5}. These p-values were obtained from the empirical distribution functions of the bootstrap sample according to Step (iii) of Algorithm \ref{alg1}. In general, we reject the null hypothesis \eqref{h0} at level $\alpha$ if the maximum of the two individual p-values for \eqref{hypothesesE} and \eqref{hypothesesT} is smaller than or equal to $\alpha$. In our example, this only holds for $\epsilon=(0.2,0.2)$ since the individual p-values are given by $\hat F_{n_{boot}}^E(\hat d^E)=0.04$ and $\hat F_{n_{boot}}^T(\hat d^T)=0.03$ such that $\max{(0.03,0.04)}=0.04 < 0.05 = \alpha$.

\section{Conclusions and discussion}
\label{sec:conc}
\def\theequation{6.\arabic{equation}}
\setcounter{equation}{0}

In the first part of this paper we investigated a single efficacy response given by a binary outcome and derived a test procedure for the equivalence of the corresponding dose-response curves, which can be modelled, for instance, by a parametric logistic regression or a probit model. We developed a parametric bootstrap test and decide for equivalence if the maximum deviation between the estimated dose response profiles is sufficiently small. 
We also considered the situation of an additional second toxicity endpoint to model the joint efficacy-toxicity responses. For this purpose we assumed efficacy and toxicity to be observed simultaneously resulting in bivariate (correlated) binary outcomes and  used a Gumbel model to fit the data. The bootstrap test was extended to this situation by combining two individual tests through the Intersection-Union-Principle.

In the second part of this paper we investigated the operating characteristics by means of an extensive simulation study. We demonstrated that the resulting procedures control their level and achieve reasonable power.
The choice of the equivalence threshold $\epsilon$ has a major impact on the performance of the test. The explicit choice has to be made on an individual basis and under consideration of clinical experts.

In certain settings the efficacy or toxicity responses are not modelled by binary outcomes, but rather by a continuous response. In case of two continuous outcomes, \cite{fedorov2007} considered normally distributed correlated responses which are dichotomized due to binary utility and the methodology proposed in this paper can be adapted to the situation considered by these authors. 
A further interesting situation occurs in case of mixed outcomes, where one of the response variables is continuous and the other a binary one. Modelling these types of responses is a challenging problem and not much work has been done on this topic in the literature. \citet{tao2013} investigated this situation by modelling these multiple endpoints by a joint model constructed with archimedean copula. An equivalence test for these types of outcomes is an interesting topic which we leave for future research.

\section{Software}
\label{sec6}
Software in the form of R code together with a sample
input data set and complete documentation is available online at \url{https://github.com/kathrinmoellenhoff/Efficacy_Toxicity}.

\bigskip

{\bf Acknowledgements} The authors gratefully acknowledge financial support by the
Collaborative Research Center ``Statistical modeling of nonlinear
dynamic processes'' (SFB 823, Teilprojekt  T1) of the German Research Foundation
(DFG).

\setlength{\bibsep}{1pt}
\begin{small}

\end{small}

\end{document}